\documentclass[twocolumn,english,pra,superscriptaddress]{revtex4-1}
\usepackage[utf8]{inputenc}
\usepackage{enumitem}
\usepackage{amsfonts}
\usepackage{amssymb}
\usepackage{color}
\usepackage{graphicx}
\usepackage{textcomp}
\usepackage{cancel}
\usepackage{dcolumn}
\usepackage{bm}
\usepackage{multirow}
\usepackage{float}
\usepackage{mathrsfs} 
\usepackage{braket}
\usepackage{amsmath}
\usepackage{subfigure}
\usepackage{epstopdf}

\usepackage[usenames,divipsnames,svgnames,table]{xcolor}
\usepackage[unicode=true,pdfusetitle,
bookmarks=true,bookmarksnumbered=false,bookmarksopen=false,
breaklinks=true,pdfborder={0 0 0},backref=false,colorlinks=true]{hyperref}
\hypersetup{linkcolor=NavyBlue,urlcolor=NavyBlue,citecolor=NavyBlue}
\usepackage{breakurl}

\begin{document}

\title{Generation and storage of spin squeezing via learning-assisted optimal control}

\author{Qing-Shou Tan}

\affiliation{Key Laboratory of Hunan Province on Information Photonics and
Freespace Optical Communication, College of Physics and Electronics,
Hunan Institute of Science and Technology, Yueyang 414000, China}

\author{Mao Zhang}
\affiliation{MOE Key Laboratory of Fundamental Physical Quantities Measurement,
Hubei Key Laboratory of Gravitation and Quantum Physics, PGMF and School of Physics,
Huazhong University of Science and Technology, Wuhan 430074, China}

\author{Yu Chen}
\affiliation{Tencent Lightspeed \& Quantum Studios, Shenzhen, China}

\author{Jie-Qiao Liao}
\affiliation{Key Laboratory of Low-Dimensional Quantum Structures and Quantum Control
of Ministry of Education, Key Laboratory for Matter Microstructure and Function of Hunan Province,
and Department of Physics and Synergetic Innovation Center for Quantum Effects and Applications,
Hunan Normal University, Changsha 410081, China}

\author{Jing Liu}
\email{liujingphys@hust.edu.cn}
\affiliation{MOE Key Laboratory of Fundamental Physical Quantities Measurement,
Hubei Key Laboratory of Gravitation and Quantum Physics, PGMF and School of Physics,
Huazhong University of Science and Technology, Wuhan 430074, China}

\begin{abstract}
The generation and storage of spin squeezing is an attractive topic in quantum metrology and the
foundations of quantum mechanics. The major models to realize the spin squeezing are the one- and
two-axis twisting models. Here, we consider a collective spin system coupled to a bosonic field,
and show that proper constant-value controls in this model can simulate the dynamical behavior
of these two models. More interestingly, a better performance of squeezing can be obtained when
the control is time-varying, which is generated via a reinforcement learning algorithm. However,
this advantage becomes limited if the collective noise is involved. To deal with it, we propose
a four-step strategy for the construction of a new type of combined controls, which include both
constant-value and time-varying controls, but performed at different time intervals. Compared to
the full time-varying controls, the combined controls not only give a comparable minimum value
of the squeezing parameter over time, but also provides a better lifetime and larger full amount
of squeezing. Moreover, the amplitude form of a combined control is simpler and more stable than
the full time-varying control. Therefore, our scheme is very promising to be applied in practice
to improve the generation and storage performance of squeezing.
\end{abstract}

\maketitle

\section{introduction}

Many advantages of quantum technology require the assistance of quantum resources. Squeezing
is such a resource~\cite{Yuen1976,Walls1994,Scully1997}. Consider a pair of canonical
observables $X$ and $Y$. Their deviations $\Delta X$ and $\Delta Y$ in a system satisfy
the Heisenberg uncertainty relation $\Delta X\Delta Y\geq |\langle [X,Y]\rangle|/2$.
The system is called squeezed when one of the deviations is less than the square root
of the bound above. A typical example is the squeezed vacuum state, in which the deviation
of the quadrature operator is squeezed. The squeezed light has been proved to be very
useful in many aspects of quantum information, especially in quantum metrology~\cite{Caves1981,Lang2013},
and it is now a promising candidate to be applied in the next-generation gravitational-wave
observatory on earth for the further improvement of the detection sensitivity~\cite{Grote2013}.
Apart from the light, the atoms can also present squeezing behaviors, known as the
spin squeezing~\cite{Kitagawa1993,Wineland1992,Wineland1994,Ma2011,Pezze2018,Cronin2009,
Sau2010,Vitagliano2011}. Similar to the squeezed light, the squeezed atoms can also
improve the measurement precision beyond the standard quantum limit~\cite{Ma2011,Pezze2018},
and more interestingly, witness the many-body entanglement~\cite{Guehne2009}.

In the early 1990th, two types of squeezing parameters for the quantification of spin squeezing
were provided by Kitagawa and Ueda~\cite{Kitagawa1993}, and Wineland et al.~\cite{Wineland1992,Wineland1994}.
Kitagawa and Ueda~\cite{Kitagawa1993} further proposed two different mechanisms, one- and two-axis
twisting models, for the generation of spin squeezed states. The one-axis twisting (OAT) model~\cite{Jin2009,
Jin2007,Bennett2013,Zhong2014,Gross2010,Riedel2010} can provide a precision limit at the scaling
$N^{-2/3}$ ($N$ is the particle number) and the two-axis twisting (TAT) model~\cite{Helmerson2001,
Zhang2003,Ng2003,Liu2011,Thomsen2002,Zhangyc2017,Groszkowski2020,Qin2020} provides a better scaling $N^{-1}$.
These advantages motivate the scientists to try to realize these models in experiments. Currently, the
OAT model can be readily obtained with the two-component Bose-Einstein condensate~\cite{Gross2010,Riedel2010}
or the nitrogen-vacancy centers~\cite{Bennett2013}, yet the TAT model is more difficult to realize
in practice. Several theoretical schemes have been proposed in recent years, such as utilizing the
Raman processes~\cite{Helmerson2001,Zhang2003} of Bose-Einstein condensate or the double well~\cite{Ng2003},
converting the OAT model into an effective TAT model~\cite{Liu2011}, phase-locked coupling between
atoms and photons~\cite{Zhangyc2017}, bosonic parametric driving~\cite{Groszkowski2020},
employing feedback in the measurement system~\cite{Thomsen2002}, and even using the week squeezing
of light~\cite{Qin2020}, Finding simple and experimental-friendly realizations of the TAT model and
searching ways to go beyond it for the generation of squeezing are still the major concerns in this field.

In this paper, we consider a general collective spin system coupled to a bosonic field via the
dispersive coupling, and propose an optimal control method for the generation and storage of spin
squeezing. Both the constant-value and time-varying controls are studied with and without noise.
The OAT and TAT models can be readily simulated by this system via proper constant-value controls.
The time-varying controls are generated via the Deep Deterministic Policy Gradient (DDPG)
algorithm~\cite{Lillicrap2015}, an advanced reinforcement learning algorithm. In recent
years, various machine learning algorithms~\cite{Mohri2018,Hafner2011,Kevine2020} have
been applied in many topics of quantum physics~\cite{Carleo2019,Krenn2020}, such as quantum
phase transitions~\cite{Nieuwenburg2017,Nieuwenburg2018,Che2020}, quantum parameter
estimation~\cite{Xu2019,Hentschel2010,Hentschel2011}, quantum speed limits~\cite{Zhang2018},
Hamiltonian learning~\cite{Wang2017} and multipartite entanglement~\cite{Krenn2016,Melnikov2018,Krenn2020}.
Aided by the deep reinforcement learning, Chen et al.~\cite{Chen2019} recently proposed a
scheme in the OAT-type model with few discrete pulses, which can obtain an enhanced amount
of squeezing close to the TAT model. In the collective spin system we consider, with the
help of time-varying controls generated by the DDPG algorithm, the performance of squeezing
goes significantly beyond the TAT model.

In practice, the collective spin system could be easily affected by the collective noise,
and it is unfortunate that the advantage of time-varying controls becomes limited when this
noise is involved. To deal with it, we propose a four-step strategy to generate a new type of
combined controls, which include both constant-value and time-varying controls, but performed
at different time intervals. The combined controls not only provide a similar maximum
squeezing compared to both the constant-value and time-varying controls, but also significantly
extend the lifetime and improve the full amount of squeezing over time. Due to the fact that
the combined controls are simpler and more stable than the full time-varying controls, it is
very promising to be applied in practical environment for the realization of an improved
performance than the TAT model on the generation and storage of squeezing.

\section{physical model and spin squeezing}

We consider a coupled atom-field system in which an ensemble of the two-level systems is coupled
to a single-mode bosonic field. The Hamiltonian of this system reads
\begin{equation}
H_0=\omega_{c}a^{\dagger}a+\omega_{z}J_{z}+g J_{x}(a^{\dagger}+a),
\label{eq:Hamiltonian}
\end{equation}
where $a$ $(a^{\dagger})$ is the annihilation (creation) operator of the field, which can be
realized by a cavity, and $J_m=\sum^{N}_{i=1}\frac{1}{2}\sigma^{(i)}_m$ is the collective angular
momentum operator with $N$ the particle number and $\sigma^{(i)}_m$ ($m=x,y,z$) the Pauli matrix
for the $i$th spin. $\omega_c$ and $\omega_z$ are the frequencies of the field and collective
system, respectively, and $g$ is the strength of the coupling. To help to generate spin squeezing,
we invoke the quantum control via the time-dependent modulation field with the Hamiltonian
\begin{equation}
H_{\mathrm{c}}(t) = \zeta(t)\nu\cos(\nu t)J_{z},
\end{equation}
where $\nu$ is the modulation frequency, $\zeta(t)$ is the amplitude. The model described by
Eq.~(\ref{eq:Hamiltonian}) can be realized with an ensemble of $^{87}$Rb atoms with up state
$|\!\!\uparrow\rangle:=|5^2S_{1/2},F=2,m_{F}=1\rangle$ and down state
$|\!\!\downarrow\rangle:=|5^2S_{1/2},F=1,m_{F}=1\rangle$ coupled to a microwave cavity
mode. The energy split between the hyperfine levels $|\!\!\uparrow\rangle$ and $|\!\!\downarrow\rangle$
is about 6.8GHz without  magnetic field. In the presence of magnetic field, the Zeeman or hyperfine
Paschen-Back shift has the same magnitude but opposite sign for the two hyperfine manifolds with
$g_{F=2}=1/2$ and $g_{F=1}=-1/2$~\cite{Steck2019}. Thus, the modulation Hamiltonian $H_c(t)$ can be
realized with a controllable magnetic field.

Due to the existence of noise on both the cavity and collective spin, the evolution of the total
density matrix $\rho$ for our model is governed by the master equation
\begin{eqnarray}
\partial_t\rho &=& -i\left[H_0+H_{\mathrm{c}}, \rho\right]
+\kappa \left(2a\rho a^{\dagger}-a^{\dagger}a\rho-\rho a^{\dagger}a \right) \nonumber \\
& & +\gamma \left(2J_z\rho J_z-J_z^2\rho-\rho J_z^2\right)
\label{eq:master_eq}
\end{eqnarray}
with $\kappa$ and $\gamma$ being the cavity loss rate and atoms dephasing rate, respectively.

To characterize the degree of spin squeezing generated in this system, we use the squeezing
parameter introduced by Kitagawa and Ueda~\cite{Kitagawa1993}
\begin{equation}
\xi^{2}=\frac{4}{N}(\Delta J_{n_{\bot}}^{2})_{\min},
\end{equation}
where $(\Delta J_{n_{\bot}}^{2})_{{\min}}$ is the minimum variance in a direction vertical
to the mean spin direction $\vec{n}_0$. A state is squeezed if $\xi^2<1$, and smaller $\xi^2$
indicates stronger squeezing. $\vec{n}_0$ in spherical coordinates is of the form
$(\sin\theta\cos\phi,\sin\theta\sin\phi,\cos\theta)$, where $\theta = \arccos\left(\langle J_z\rangle
/\sqrt{\langle J_x\rangle^2+\langle J_y\rangle^2+\langle J_z\rangle^2}\right)$ and
$\phi=\arccos\left({\langle J_{x}\rangle}/{\sqrt{\langle J_{x}\rangle^{2}+\langle J_{y}\rangle^{2}}}\right)$
are polar and azimuthal angles, respectively. The other two orthogonal vectors with respect to
$\vec{n}_{0}$ are $\vec{n}_{1}= (-\sin\phi, \cos\phi, 0)$ and $\vec{n}_{2} = (-\cos\theta\cos\phi,
-\cos\theta\sin\phi, \sin\theta)$. Define $J_{\vec{n}}=\vec{n}\cdot\vec{J}$ with $\vec{J}=(J_x,J_y,J_z)$,
the variance $(\Delta J_{n_{\bot}}^{2})_{{\rm min}}$ can be calculated via the equation
\begin{eqnarray}
(\Delta J_{n_{\bot}}^{2})_{{\rm min}}=\frac{1}{2}\left(\mathcal{C}-\sqrt{\mathcal{A}^{2}
+\mathcal{B}^{2}}\right),
\end{eqnarray}
where $\mathcal{C}=\langle J_{\vec{n}_{1}}^{2}+J_{\vec{n}_{2}}^{2}\rangle$, $\mathcal{A}=\langle
J_{\vec{n}_{1}}^{2}-J_{\vec{n}_{2}}^{2}\rangle$, and $\mathcal{B}=\langle J_{\vec{n}_{1}}J_{\vec{n}_{2}}
+J_{\vec{n}_{2}}J_{\vec{n}_{1}}\rangle$. Then the squeezing parameter becomes
\begin{equation}
\xi^{2}=\frac{2}{N}\left(\mathcal{C}-\sqrt{\mathcal{A}^{2}
+\mathcal{B}^{2}}\right).
\end{equation}
The corresponding optimal squeezing direction is
\begin{equation}
\vec{n}_{\mathrm{opt}}=\vec{n}_1\cos\left(\varphi_{\mathrm{opt}}\right)
+\vec{n}_2\sin\left(\varphi_{\mathrm{opt}}\right),
\end{equation}
where the optimal angle $\varphi_{\mathrm{opt}}$ reads~\cite{Ma2011}
\begin{equation}
\varphi_{\mathrm{opt}}=\begin{cases}
\frac{1}{2}\arccos\left(\frac{-\mathcal{A}}{\sqrt{\mathcal{A}^2+\mathcal{B}^2}}\right),
~\mathrm{for}~\mathcal{B}\leq 0, \\
\pi-\frac{1}{2}\arccos\left(\frac{-\mathcal{A}}{\sqrt{\mathcal{A}^2+\mathcal{B}^2}}\right),
~\mathrm{for}~\mathcal{B}>0.
\end{cases}
\end{equation}

In our scheme, we assume that the ensemble of atoms is prepared in a coherent spin state, which is
defined as
\begin{equation}
\left|\eta\right\rangle=(1+\left|\eta\right|^{2})^{-J}\sum_{m=-J}^{J}\left(\begin{array}{c}
2J\\
J+m
\end{array}\right)^{1/2}\eta^{J+m}\left|J,m\right\rangle,
\end{equation}
where $\left|J,m\right\rangle$ ($J=N/2$, $m=0, \pm 1,\cdots,\pm J$ for an even $N$ and
$\pm\frac{1}{2},\cdots,\pm J$ for an odd $N$) is known as the Dicke state, namely, the eigenstate
of $J_{z}$ with eigenvalue $m$. Since $\eta$ can be expressed by $\eta =-\tan(\frac{\theta}{2})
\exp(-i\phi)$, the coherent spin state can also be written as $|\theta,\phi\rangle$. In the
following, the initial state of the atoms are taken as the coherent spin state
$\left|\frac{\pi}{2},\frac{\pi}{2}\right\rangle$ and the initial state of the cavity is
the vacuum state.

\section{Control-enhanced spin squeezing}

\subsection{Constant-value control}

\begin{figure}[tp]
\centering\includegraphics[width=8cm]{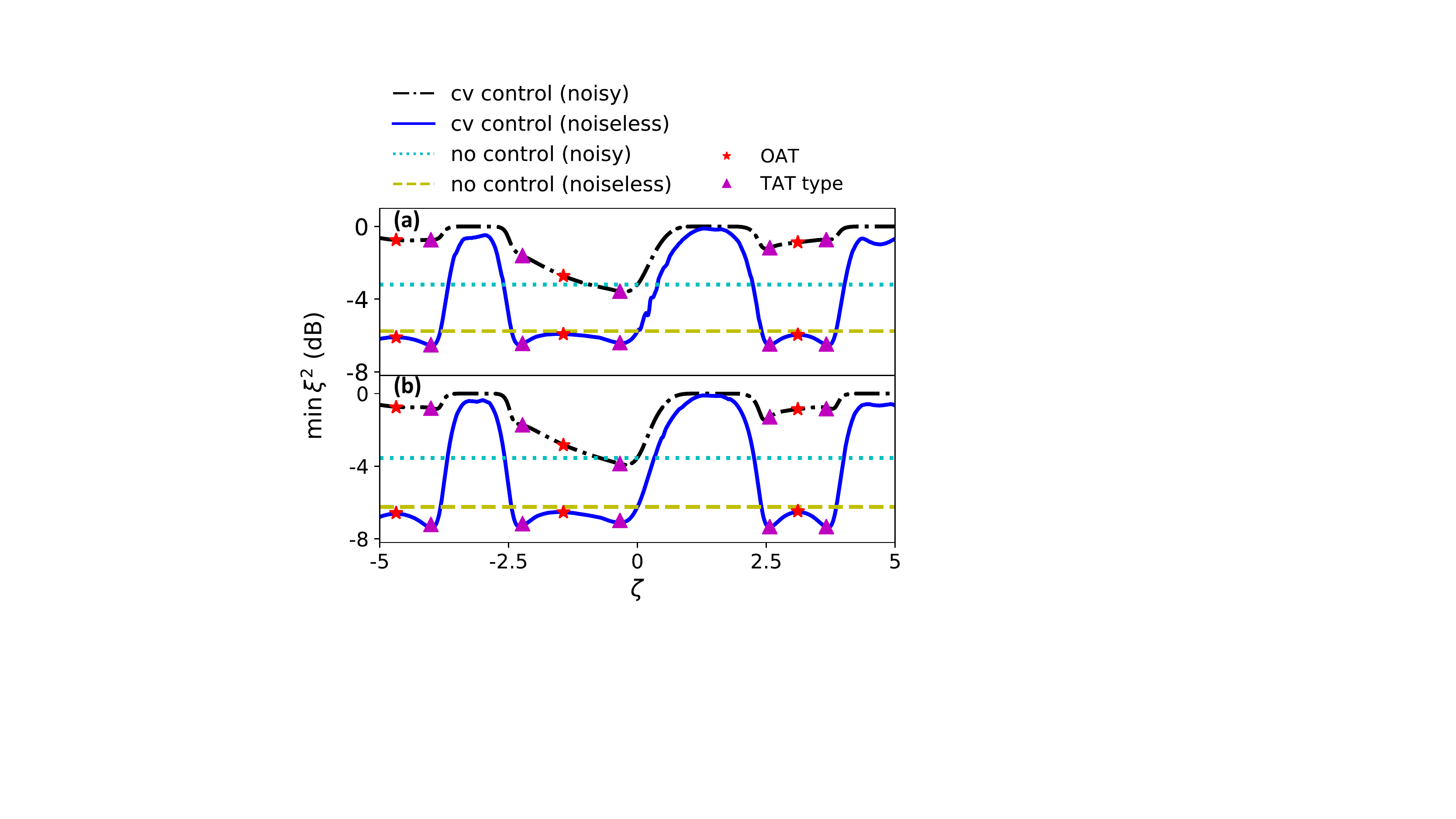}
\caption{The minimum value of $\xi^2$ ($\min\,\xi^2$) as a
function of $\zeta$ in the case of (a) $N=6$ and (b) $N=8$ for
both noise and noiseless scenarios. The dash-dotted black and solid blue
lines represent $\min\,\xi^2$ for the constant-value (cv) controls
with and without noise. The dotted cyan and dashed yellow lines
represent $\min\,\xi^2$ in the non-controlled scenario with and
without noise. The values of $\min\,\xi^2$ for the effective OAT
and TAT-type Hamiltonians are shown as the red stars and purple triangles.
The decay rates are set to be $\kappa=\gamma=0.01g$ in the plots.}
\label{fig:cv}
\end{figure}

The constant-value control refers to invoking a time-independent value of control amplitude, i.e.,
$\zeta(t)=c$. This control is simple, economic and easy to be implemented in experiments. The
Hamiltonian in Eq.~(\ref{eq:Hamiltonian}) with a proper constant-value control can simulate a OAT
or TAT-type Hamiltonian. In the rotating frame defined by $V(t)=\mathcal{T}\exp\left[i\int_{0}^{t}(H_{0}
+H_{\mathrm{c}})d\tau\right]$ ($\mathcal{T}$ is the time-ordering operator), the Hamiltonian can
be approximately rewritten into
\begin{equation}
\tilde{H}\approx(g_{0}J_{+}a e^{-i\delta t}+g_{m_0}J_{-}a e^{-i\Delta_{m_{0}}t})+\mathrm{H.c.},
\label{eq:eff_H}
\end{equation}
when the condition $\nu\gg g, \delta$ is satisfied. Here $J_{\pm}=J_x \pm iJ_y$, $g_{0}=g\mathcal{J}_{0}(\zeta)/2$,
$g_{m_0}=g\mathcal{J}_{m_{0}}(\zeta)/2$, $\delta = \omega_c-\omega_z$, and $\Delta_{m_{0}}=m_{0}\nu
+\omega_{c}+\omega_{z}$ where $m_0$ is the optimal integer $m$ to reach the minimum value of
$|m\nu+\omega_c+\omega_z|$. $\mathcal{J}_{n}(\zeta)$ is the $n$th Bessel function of the first
kind. H.c. represents the Hermitian conjugate. In the case that $\delta =\Delta_{m_{0}} \equiv \Delta$,
namely, $m_{0}=-2[\omega_{z}/\nu]$ ($[\cdot]$ is the rounding function), and rotating $\tilde{H}$ back
to a non-rotating frame with $V_1=\exp(i\Delta a^{\dagger}at)$, an effective Hamiltonian
$H_1=\Delta a^{\dagger}a+g(\Sigma^{\dagger}a+\Sigma a^{\dagger})$ can be obtained with
$\Sigma=[\mathcal{J}_{0}(\zeta){J}_{-}+\mathcal{J}_{m_{0}}(\zeta){J}_{+}]/2$.

For large detunings $\Delta\gg g > g_{0}\,(g_{m_0})$, taking the transformation
$e^{R}H_1e^{-R}$ with $R = \frac{g}{\Delta}(\Sigma a^{\dagger}-\Sigma^{\dagger}a)$, and truncating to
the second order of $g/\Delta$, the effective Hamiltonian becomes (the details can be found in the Appendix)
\begin{eqnarray}
H_{\rm eff} &=& \Delta a^{\dagger}a-\frac{g_{0}^{2}-g_{m_0}^{2}}{\Delta}(1+2a^{\dagger}a)J_{z} \nonumber \\
& & +\frac{\left(g_{0}-g_{m_0}\right)^{2}}{\Delta}J_{z}^{2}-\frac{4g_{0}g_{m_0}}{\Delta}J_{x}^{2}.
\label{eq:H_eff_diag}
\end{eqnarray}
An effective OAT model $\chi J_{x}^{2}$ can be obtained by the equation above when taking $g_{0}=g_{m_0}$
and the coefficient $\chi=-g^{2}\mathcal{J}_{m_{0}}^{2}(\zeta)/\Delta$. It is easy to see that when no
control is involved ($\zeta=0$), the transformed Hamiltonian is also an OAT-type (with linear term)
model~\cite{Bennett2013} since $\mathcal{J}_{0}(0)=1$ and $\mathcal{J}_{m\neq 0}(0)=0$.
Some values for $\zeta$ to simulate the OAT model are $-4.680, -1.435$, and $3.113$ for $m_0=-1$. Moreover,
Eq.~(\ref{eq:H_eff_diag}) can also simulate the TAT-type model
\begin{eqnarray}
H_{\rm TAT-type}=-\lambda(J_{x}^{2}-J_{z}^{2})-\lambda^{\prime}(1+2a^{\dagger}a)J_{z},
\end{eqnarray}
when the coefficients satisfying $(g_{0}-g_{m_0})^{2}=4g_{0}g_{m_0}$, and the coefficients
$\lambda=4g_0 g_{m_0}/\Delta$ and $\lambda^{\prime}=(g_0^2-g_{m_0}^2)/\Delta$. Some values for
$\zeta$ to satisfy this condition are $-2.284, -0.338$, and $2.569$ for $m_0=-1$. The linear term
can be eliminated on average through a dynamical decoupling protocol, which makes the transformed
Hamiltonian a standard TAT Hamiltonian. The validity of this effective Hamiltonian are checked by
calculating the fidelity between the evolved states given by the original and effective Hamiltonians,
which is shown in the Appendix.

Apart from simulating the OAT and TAT models, searching ways to go beyond these models on the generation
of squeezing is also crucial. Hence, the performance of a general constant-value control is studied
for both noisy and noiseless scenarios, as shown in Fig.~\ref{fig:cv}(a) for $N=6$ and Fig.~\ref{fig:cv}(b)
for $N=8$. The parameters are set as $w_z=110, w_c=100, \nu=200$ and $g=1$. In the noiseless scenario,
the minimum values of $\xi^2(t)$ (solid blue lines, denoted by $\min \xi^2$), show a large amplitude
waving behavior when the control amplitude $\zeta$ varies, and not all the values are capable to provide
an enhanced squeezing than the non-controlled one ($\zeta=0$, dashed yellow lines). The TAT-type
Hamiltonians provide a good squeezing performance (purple triangles) in the regime ([-5,5]) given in
the plot. The corresponding values of $\min \xi^2$ are very close to the optimal ones. For a larger
regime (for example [-10, 10]), the optimal values of $\zeta$ can provide a better performance than
the TAT-type Hamiltonians, yet they may also have to face the difficulty of generation in practice.
The OAT Hamiltonians (red stars) do not present an obvious advantage on squeezing than the non-controlled
one in our case.

In the case of involving collective noise, the performance of constant-value controls deteriorates
significantly. The maximum squeezing given by the constant-value controls (dash-dotted black lines)
can only surpass the non-controlled one (dotted cyan lines) in a very narrow regime. Hence, with the
existence of collective noise, the enhancement of squeezing given by the constant-value controls could
be very limited, and if it is the only choice, then a better strategy is let the system evolves freely.

\subsection{Time-varying control}

\begin{figure}[tp]
\centering\includegraphics[width=8.5cm]{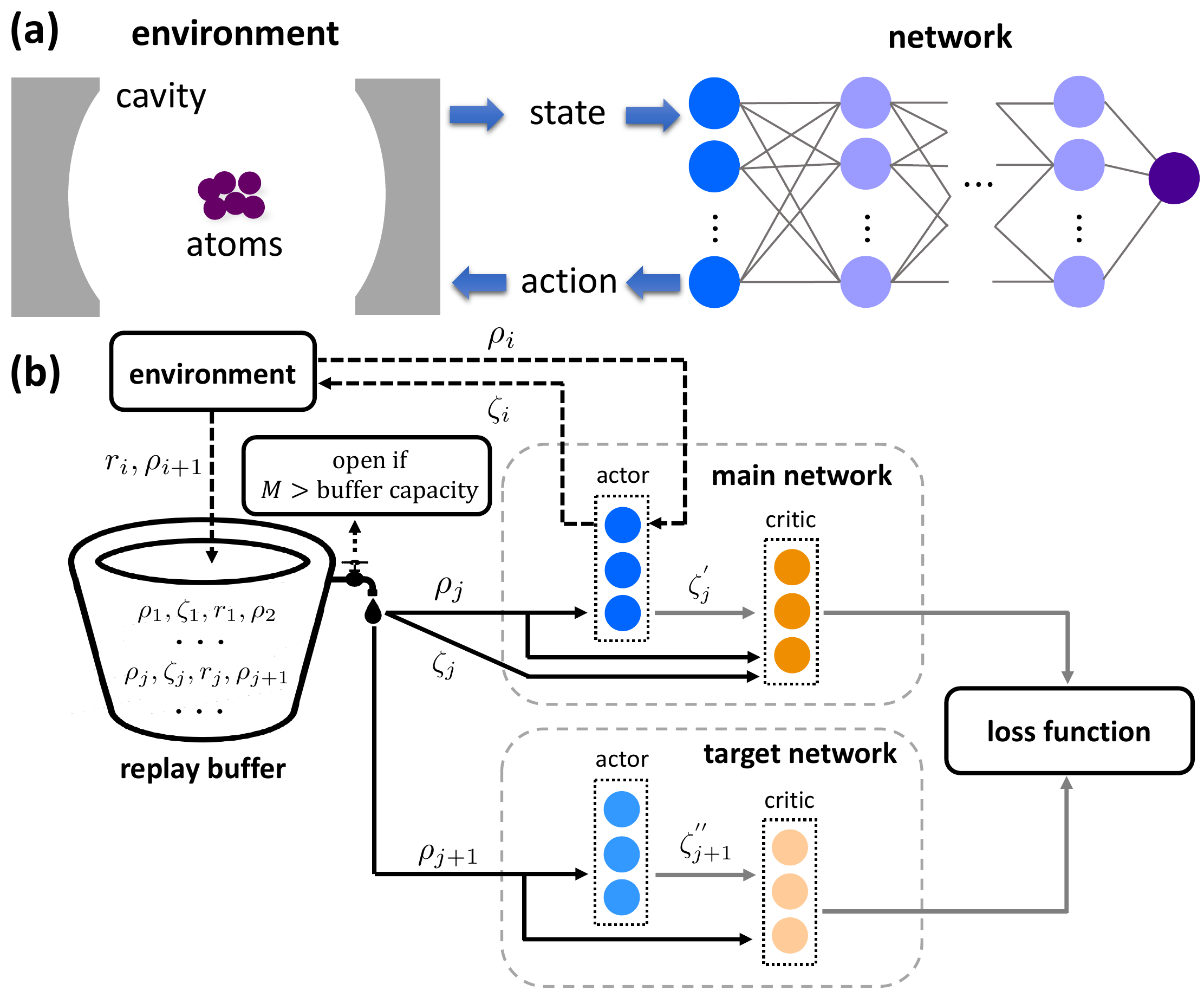}
\caption{(a) Schematic of control generation for the enhancement
of spin squeezing via reinforcement learning in a collective
spin-half system coupled to a cavity. (b) Brief flow chart of
the DDPG algorithm~\cite{Lillicrap2015}.}
\label{fig:setup}
\end{figure}

In many scenarios, time-varying controls ($\zeta=\zeta(t)$) are more powerful than the constant-value
controls since they provide a way larger parameter space for the control. Finding an optimal
time-varying control for a given target is a major concern in quantum control. Many algorithms,
including the GRAPE~\cite{Khaneja2005,Fouquieres2011,Liu2017a,Liu2017b,Liu2020,Wu2019,Peng2018},
krotov's method~\cite{Reich2012,Goerz2019}, and machine learning~\cite{Xu2019,Fosel2018,Bukov2018,
An2019,Yang2020,Bukov2018} have been employed into various scenarios in quantum physics, like quantum
information processing and parameter estimation, for the generation of optimal control. With respect
to the spin squeezing, Pichler et al.~\cite{Pichler2016} recently used the chopped random basis
technique in the OAT model and obtained an enhanced behavior of squeezing than the adiabatic
evolution. In the following we employ the DDPG algorithm~\cite{Lillicrap2015}, an advanced
reinforcement learning algorithm to study the performance of time-varying controls on the
generation and storage of spin squeezing.

\begin{figure}[tp]
\centering\includegraphics[width=8.5cm]{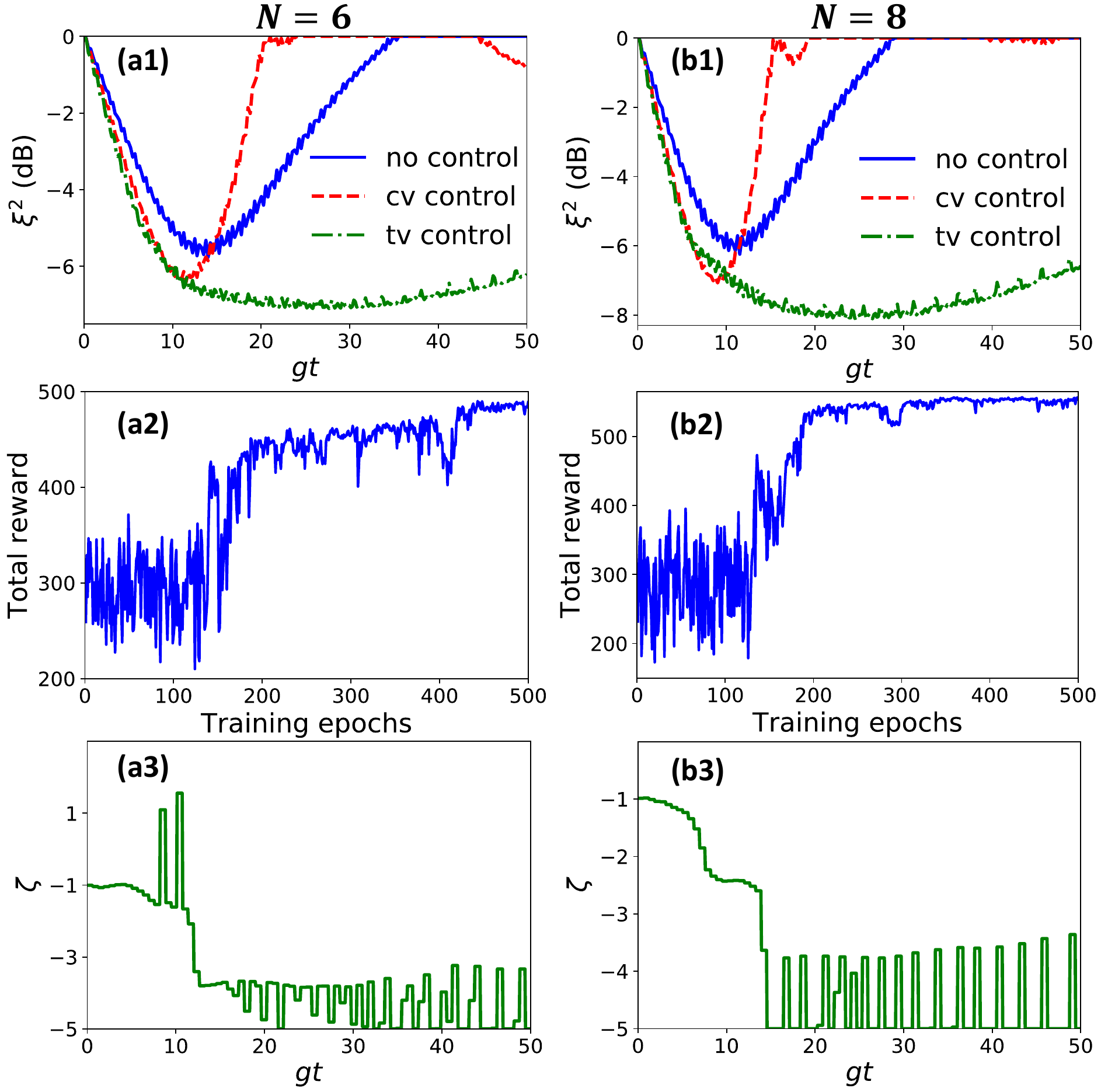}
\caption{The results for unitary dynamics in the case of $N=6$
(a1-a3) and $N=8$ (b1-b3). Panels (a1) and (b1) show the evolution
of $\xi^2$ (in the unit of dB) with the time-varying (tv) control
(dash-dotted green lines), optimal constant-value (cv) control
(dashed red lines) and without control (solid blue lines), respectively.
The time-varying controls are generated via the learning with the
corresponding total reward given in (a2) and (b2). Panels (a3) and
(b3) give the optimal control $\zeta$ with respect to the dash-dotted
green lines in (a1) and (b1).}
\label{fig:unitary}
\end{figure}

Reinforcement learning uses a network (also called agent) to provide choices of actions for
the environment to improve the reward. Its process in our case is illustrated in
Fig.~\ref{fig:setup}(a). The environment, consisting of the cavity and atoms, generates a
dynamical quantum state according to Eq.~(\ref{eq:master_eq}), then sends it to the network,
which provides an action (the control amplitude $\zeta$) accordingly. Next, the environment
uses this action to evolve and generates a new state, and again sends it to the network for
the generation of next control. A typical algorithm of the reinforcement learning is the
Actor-Critic algorithm, in which the critic network is used to evaluate the reward that the
actor obtained. The DDPG algorithm is an advanced Actor-Critic algorithm, and includes a
replay buffer and additional two target networks besides the main actor and critic networks,
as shown in Fig.~\ref{fig:setup}(b). In the first $M$ epochs of an episode, the control
amplitude is generated randomly via the main actor network and all the data of $\rho_j$
(density matrix), $\zeta_j$ (control amplitude), $r_j$ (reward) and $\rho_{j+1}$ are saved
in the buffer (the black barrel in Fig.~\ref{fig:setup}(b)). The subscript $j$ and $j+1$
represent the $j$th and $(j+1)$th time steps. The reward in our case is taken as
$r_j = -10\mathrm {log}_{10}(\xi^2_j)$ with $\xi^2_j$ the squeezing parameter at the $j$th
time step. Beyond the $M$th epoch, the buffer picks a random array of data $(\rho_j,\zeta_j,
r_j,\rho_{j+1})$ and sends $(\rho_j,\zeta_j)$ to the main network and $\rho_{j+1}$ to the
target network. The outputs of both networks construct the loss function, which is used
to update the main networks. The target networks update much slower than the main networks
since they only absorb a small weight (such as $10\%$) of the main networks.

\begin{figure}[tp]
\centering\includegraphics[width=8cm]{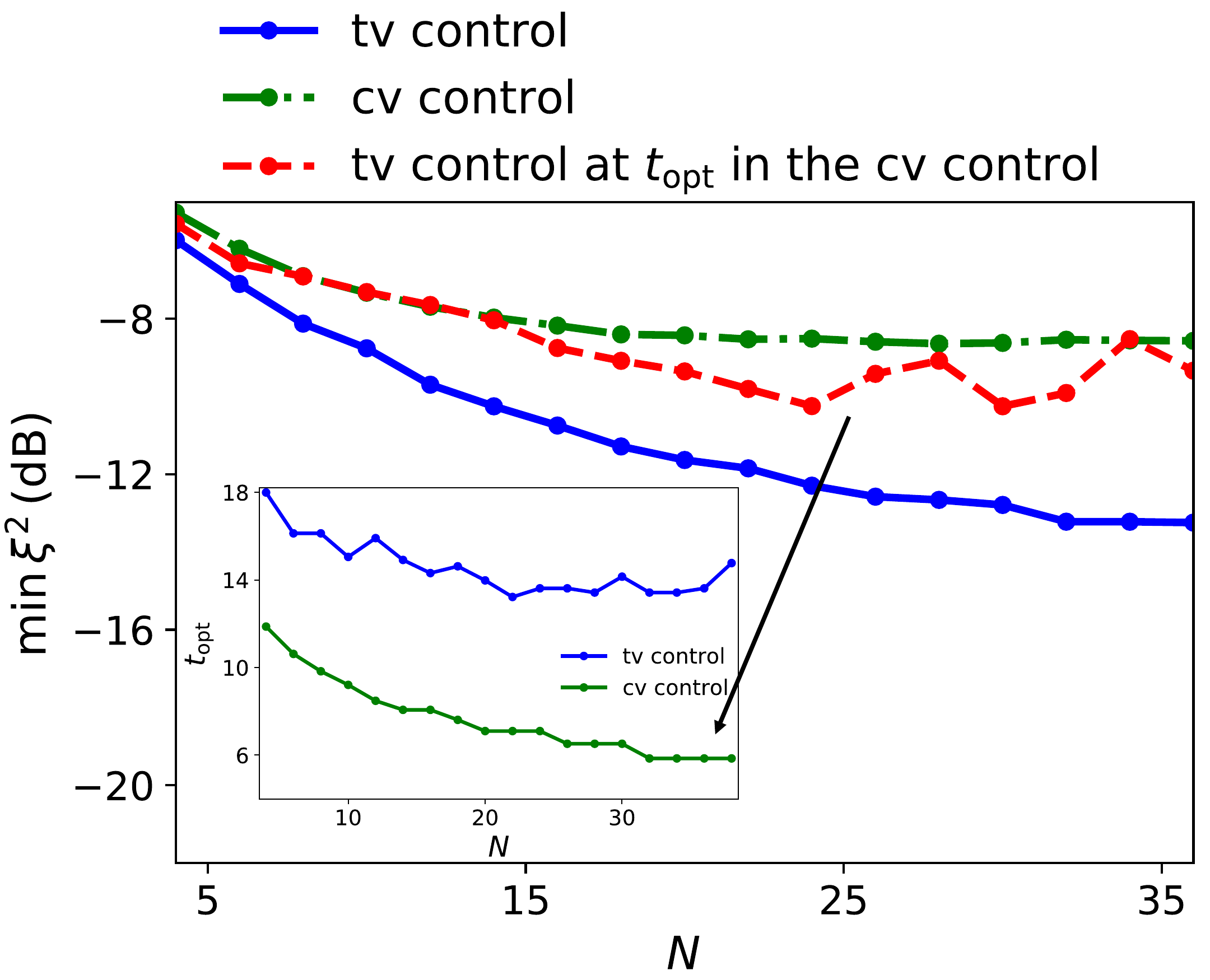}
\caption{The variety of spin squeezing as a function of particle
number $N$ for the unitary dynamics. The dash-dotted green and
solid blue lines represent the minimum spin squeezing with the
optimal constant-value control and time-varying control. The
dashed red line represents the value of spin squeezing with
the time-varying control at the time that achieve the minimum
squeezing in the optimal constant-value control. The parameters
are set to be the same with Fig.~\ref{fig:unitary}. The green and
blue lines in the insert are the optimal time for the minimum
squeezing in the case of optimal constant-value and time-varying
controls, respectively.}
\label{fig:xi_N}
\end{figure}

The performance of the time-varying control generated via the DDPG for the noiseless case
is shown in Fig.~\ref{fig:unitary}(a1) for $N=6$ and Fig.~\ref{fig:unitary}(b1) for $N=8$.
The squeezing parameter $\xi^2$ (in the unit of dB) given by the time-varying control
in both cases (dash-dotted green lines) are significantly lower than those with the
optimal constant-value control (dashed red lines) and without control (solid blue lines)
for almost all the time. With the time-varying control, not only the minimum value of
$\xi^2(t)$ is lower, but $\xi^2(t)$ also keeps in a low position for a significant long
time. The DDPG algorithm works well in our case as the total reward converges with around
500 training epochs, as shown in Figs.~\ref{fig:unitary}(a2) and (b2). The corresponding
optimal control is illustrated in Figs.~\ref{fig:unitary}(a3) and (b3). Although the
optimal constant-value control can provide a lower minimum value of $\xi^2$ than the
non-controlled case, $\xi^2$ grows very fast afterwards, indicating a shortage on the
storage of squeezing.

The behavior of spin squeezing with the increase of particle number is also a major
concern in the study of this field, which is provided in Fig.~\ref{fig:xi_N}. It
shows that the advantage of time-varying control (solid blue line) on the minimum
squeezing enlarges with the increase of $N$ compared to the optimal constant-value
control (dash-dotted green line). The trade-off is that the corresponding evolution
time is longer as given in the insert. More interestingly, the performance of
time-varying control (dashed red line) at the time (green line in the insert)
to achieve the minimum squeezing in the case of optimal constant-value control
also goes beyond the optimal constant-value control for most values of $N$ in the
plot, especially around $N=20$ to $30$, indicating that time-varying control can
still present a better behavior for a short time-scale in this regime.

\begin{figure}[tp]
\centering\includegraphics[width=8.5cm]{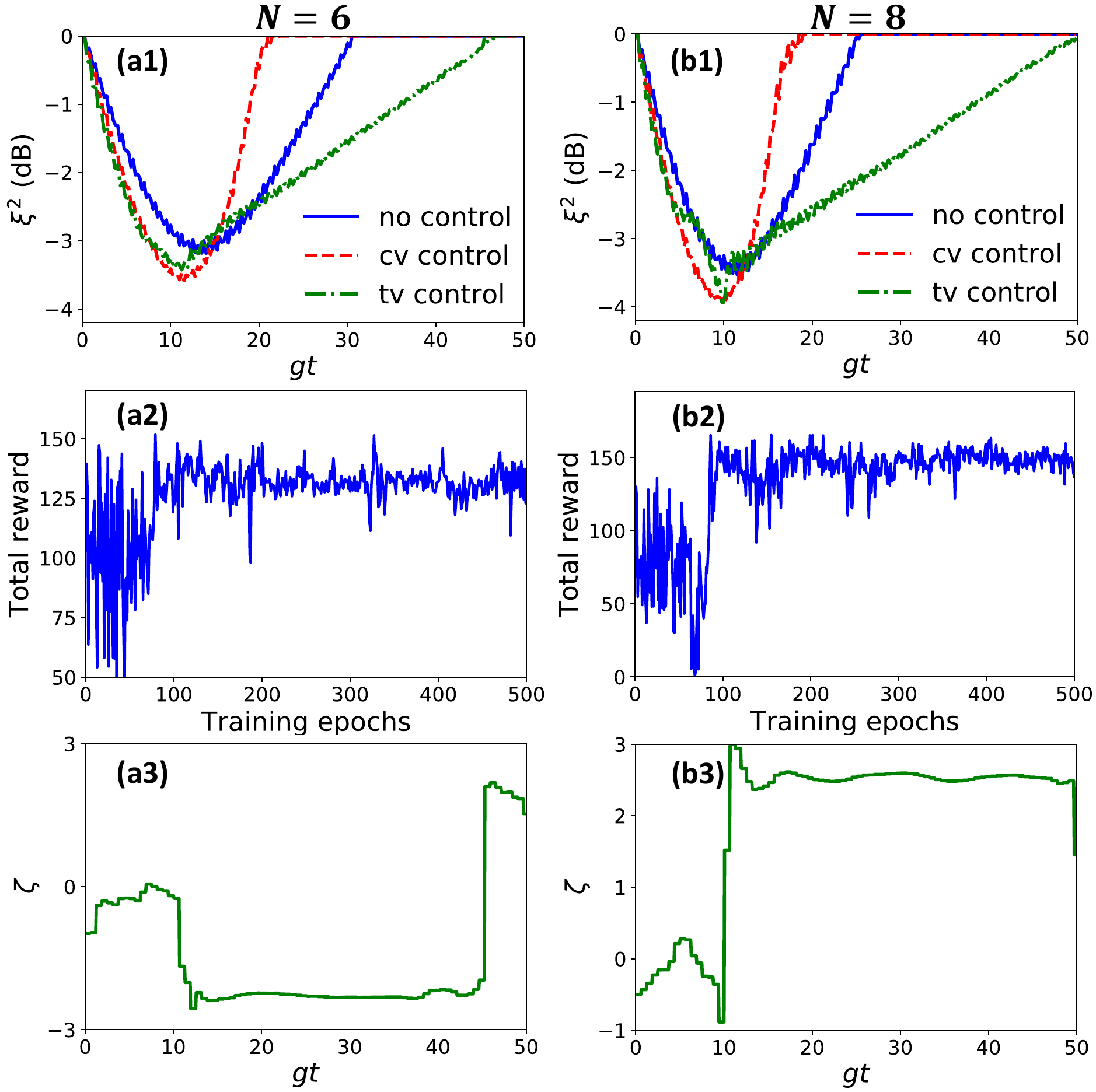}
\caption{The results for noisy dynamics in the case of $N=6$ (a1-a3) and
$N=8$ (b1-b3). Panels (a1) and (b1) show the evolution of $\xi^2$ (in the
unit of dB) with the time-varying (tv) control (dash-dotted green lines),
optimal constant-value (cv) control (dashed red lines) and without control
(solid blue lines), respectively. The time-varying controls are generated
via the learning with the corresponding total rewards given in (a2) and (b2).
Panels (a3) and (b3) give the optimal control $\zeta(t)$ with respect to the
dash-dotted green lines in (a1) and (b1). The noisy dynamics in the plots
are governed by Eq.~(\ref{eq:master_eq}) with the decay rates
$\kappa=\gamma=0.01g$.}
\label{fig:noisy}
\end{figure}

\begin{figure}[tp]
\centering\includegraphics[width=8cm]{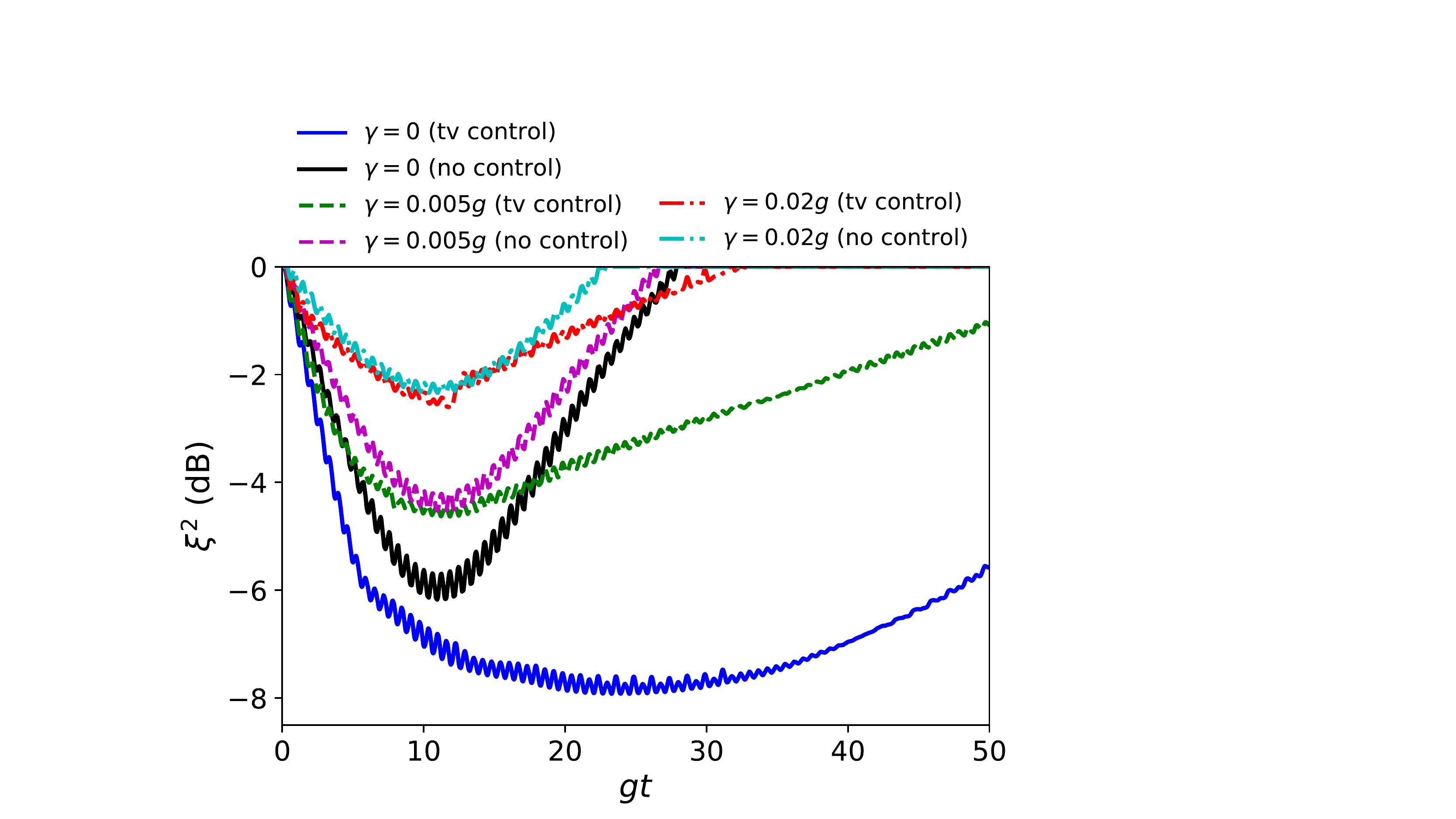}
\caption{The evolution of $\xi^2$ with different dephasing rate
with time-varying control or without control in the case of $N=8$.
Other parameteres are set to be the same as Fig.~\ref{fig:noisy}.}
\label{fig:dephasing_rate}
\end{figure}

Taking into account the collective noise ($\gamma=0.01g$), the advantage of time-varying
control in the unitary dynamics becomes limited. The minimum value basically coincides with
the one from the optimal constant-value control, as shown in Fig.~\ref{fig:noisy}(a1) for
$N=6$ and Fig.~\ref{fig:noisy}(b1) for $N=8$. The corresponding total rewards and control
amplitudes are given in Figs.~\ref{fig:noisy}(a2), (b2), (a3) and (b3), respectively.
With respect to the storage of squeezing, we first define the integral
\begin{equation}
S=-10\int^{\infty}_{0}\log_{10}\xi^{2}(t) dt,  \label{eq:S}
\end{equation}
the full amount of squeezing (dB) that the system generates over all time as the
quantification of the storage of squeezing. Now denote $S_{\mathrm{tv}}$, $S_{\mathrm{cv}}$,
and $S_{\mathrm{no}}$ as the storage of squeezing in the noisy case with the time-varying control,
optimal constant-value control (corresponds to the minimum $\min\xi^2$) and without control,
respectively. In Fig.~\ref{fig:noisy}, for $N=6$, we have $S_{\mathrm{tv}}\approx85.76$,
$S_{\mathrm{cv}}\approx 46.76$, and $S_{\mathrm{no}}\approx 57.90$. These values become
$S_{\mathrm{tv}}\approx 96.06$, $S_{\mathrm{cv}}\approx 43.02$, and $S_{\mathrm{no}}\approx 52.61$
for $N=8$. One can see that $S_{\mathrm{cv}}$ is less than $S_{\mathrm{no}}$ in both cases, which
means under the collective noise, the performance of the optimal constant-value control on the
storage of squeezing is worse than that without control. The constant-value control, including
the TAT-type model, is not the optimal choice from the aspect of storage. Although the time-varying
control does not present an obvious advantage on the minimum value of $\xi^2$, its capability of
storage still significantly outperforms the other schemes, and it grows with the increase of particle
number. Moreover, when the dephasing rate is larger, although the performance of time-varying control
deteriorates, as shown in Fig.~\ref{fig:dephasing_rate}, but it is still significantly
better than the non-controlled case, which shows the power of control in this model.

From the analysis above, one may notice that the constant-value controls can provide a good
minimum value of $\xi^2$ and the time-varying controls show an advantage on the storage of
squeezing. To combine the advantages of both the constant-value and time-varying controls,
in the following we propose a four-step strategy to generate a new type of combined controls
for the enhanced generation and storage of squeezing.

\subsection{Combined control}

\begin{figure*}[tp]
\centering\includegraphics[width=17cm]{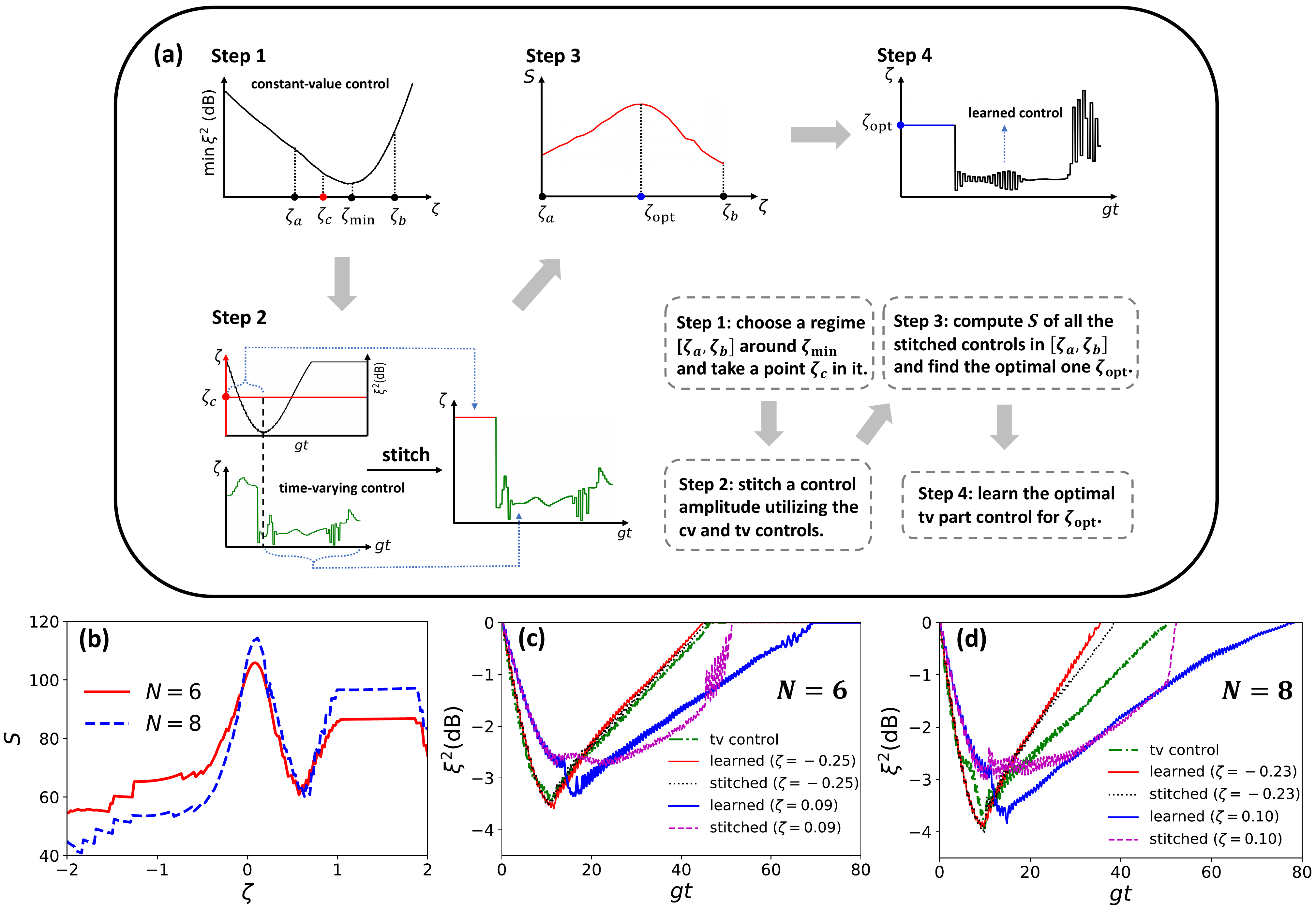}
\caption{The illustration of the combined strategy and the corresponding results.
(a) Four steps for the generation of a combined control. Step 1: choose a reasonable
regime $[\zeta_a, \zeta_b]$ around the optimal point $\zeta_{\min}$ that gives the
minimum value of $\min\xi^2$ in the scenario of constant-value controls. Step 2:
construct the stitched controls with the constant-value part given in $[\zeta_a, \zeta_b]$.
Step 3: calculate $S$ for all stitched controls and find the optimal point
$\zeta_{\mathrm{opt}}$ that gives the maximum $S$. Step 4: learn the time-varying
part of the $\zeta_{\mathrm{opt}}$'s stitched control and construct the final
combined control. (b) $\mathcal{S}$ as a function of the stitched controls. This
plot is used to search $\zeta_{\mathrm{opt}}$. (c-d) The comparison of $\xi^2(t)$
given by different controls for $N=6$ and $N=8$. The parameters are set the same
as those in previous figures.}
\label{fig:combine}
\end{figure*}

The proposed strategy consists of both constant-value and time-varying controls that performed
at different time intervals and aims at improving the storage of squeezing. We first show how
to use this strategy to generate a combined control. There are four steps to perform this strategy,
as given in Fig.~\ref{fig:combine}(a). The first step is to find the optimal constant-value control
$\zeta_{\min}$, which corresponds to the minimum value of $\min \xi^2$ and then choose a reasonable
regime $[\zeta_a, \zeta_b]$ around $\zeta_{\min}$. Next, as the second step, we stitch a control
amplitude with any value ($\zeta_c$) in $[\zeta_a, \zeta_b]$ and the previous learned full time-varying
control in this case. Specifically, denote the time that $\xi^2(t)$ reaches its minimum value under
the constant-value control $\zeta_c$ as $t_{\min}$, then the stitched control in the time interval
$[0,t_{\min}]$ is the constant value $\zeta_{c}$, and after $t_{\min}$, the control amplitude copies
the corresponding part of the full time-varying control, as illustrated in Step 2 in Fig.~\ref{fig:combine}(a).
The third step is to calculate $S$ in Eq.~(\ref{eq:S}) for all the stitched controls generated from
all points in the regime $[\zeta_a, \zeta_b]$ and find the optimal one ($\zeta_{\mathrm{opt}}$)
which gives the maximum value of $S$. The last step is to replace the time-varying part of the
stitched control of $\zeta_{\mathrm{opt}}$ to an optimal one that the DDPG algorithm finds. This
is the final combined control.

In this strategy, the reward is still taken as the squeezing parameter, i.e.,
$r_j = -10\mathrm {log}_{10}(\xi^2_j)$. In the DDPG algorithm, the actual target for the
agent to maximize at $j$th time step is the discount reward $R_j=\sum^k_{j^{\prime}=j}
\mu^{j^{\prime}-j}r_{j^{\prime}}$ with $\mu\in(0,1]$ the discount factor and $k$ the number
of time steps. The design of discount reward is to reflect the variety of influence for the
control at $j$th time step on the rewards afterwards, which reduces with the increase of time
intervals. As the discount reward contains the information of rewards for all time afterwards,
it basically has a positive correlation with $S$ defined in Eq.~(\ref{eq:S}). Hence, we can
still use this reward form for the optimization of the storage of squeezing.

In many realistic cases, finding all the learned time-varying parts in the entire regime
$[\zeta_a, \zeta_b]$ for the calculation of $S$ could be very time-consuming, this is the
reason why we need to construct the stitched controls in the second step. The stitched control
only requires an episode (500 epochs in our case) of Learning to find an optimal full time-varying
control. Different constant-value parts may have different time [$t_{\min}(\zeta)$] to reach
the minimum value of $\min \xi^2$, hence, one may need to truncate different parts of the full
time-varying control for the further construction of the stitched controls.

Utilizing this strategy, the behaviors of $\mathcal{S}$ in our model are shown in
Fig.~\ref{fig:combine}(b) for both $N=6$ (solid red line) and $N=8$ (dashed blue line).
The maximum values of $S$ are saturated by very small values of $\zeta$, namely, $\zeta_{\mathrm{opt}}
\approx 0.09$ for $N=6$ and $0.10$ for $N=8$. The performances of both stitched and combined
(learned) controls that final obtained are given in Figs.~\ref{fig:combine}(c), and \ref{fig:combine}(d)
for $N=6$ and $N=8$, respectively. In the case of $N=6$, the performances of the stitched control
(dotted black line) and learned control (solid red line) with the $\zeta_{\min}$ point ($\zeta=-0.25$)
as the constant-value parts basically coincide with each other, and also coincide with the
performance given by the full time-varying control (dash-dotted green line). Hence, one does not
need a full time-varying control here for the generation and storage of squeezing, the stitched
or the learned control can realize the same performance but with a more simple and stable control
amplitude due to the fact that both these controls have a constant-value part. In the case of $N=8$,
the stitched and learned controls with the $\zeta_{\min}$ point ($\zeta=-0.23$) also coincide
with each other, however, different with the case of $N=6$, they are worse than that given by
the full time-varying control, which indicates that it is not a good choice to use $\xi_{\min}$
for the construction of the combined control.

Varying from the phenomenon with the $\zeta_{\min}$ points, the stitched and final combined
controls with $\zeta_{\mathrm{opt}}$ as the constant-value part show very different behaviors,
which supports our strategy that $\zeta_{\mathrm{opt}}$, rather than $\zeta_{\min}$, should be
used for the construction of final combined controls. The squeezing parameters with the stitched
controls (dashed purple lines in Figs.~\ref{fig:combine}(c-d)) have a bad minimum value compared
to the full time-varying controls and grows very fast after the time around $gt=50$, which may
be due to the fact the time-varying part of this control comes from the full time-varying control,
and this part becomes dominant with the passage of time. However, the final (learned) combined
controls show a very good performance. As a matter of fact, the existence of squeezing with the
combined controls lasts much longer than those with other controls. The squeezing parameters
for the combined controls (solid blue lines in Figs.~\ref{fig:combine}(c-d)) vanish at
around $gt=70$ for $N=6$ and $gt=80$ for $N=8$. In the meantime, the full time-varying controls
can only provide a lifetime of squeezing within or around $gt=50$ in both cases. Moreover, the
storage of squeezing with the combined controls also outperforms that with the time-varying controls.
Quantitatively to say, $S_{\mathrm{c}}$ ($S$ for the combined controls) is about $117.60$ for $N=6$
and $129.76$ for $N=8$, which increases about $35\%\sim 40\%$ compared to $S_{\mathrm{tv}}$.
Besides, the minimum values of $\xi^2(t)$ are very close to those with the time-varying controls
in both cases.

\begin{figure}[tp]
\centering\includegraphics[width=8cm]{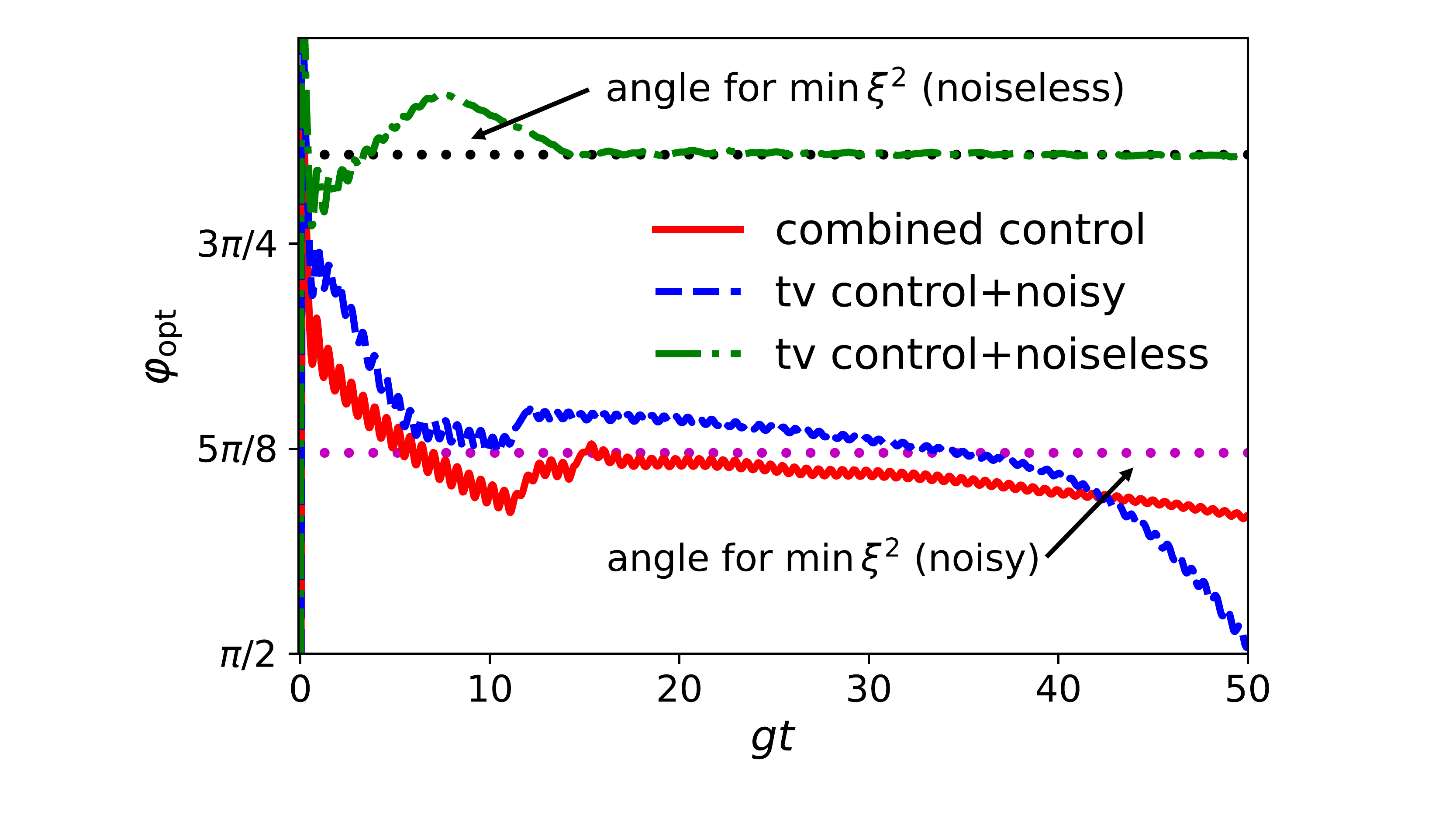}
\caption{The evolution of optimal squeezing angle in the case
of the time-varying control under noiseless dyanmics (dash-dotted
green line) and noisy dynamics (dashed blue line), and in the case
of the combined control (solid red line). The dotted pruple and
black lines represent the angles for $\min \xi^2$ under noisy
and noiseless dynamics. The decay rates for the blue and red
lines are set to be $\kappa=\gamma=0.01g$. Other parameters are
set to be same with the previous figures.}
\label{fig:angle}
\end{figure}

In physics, an intuitive picture for the storage of squeezing with control is to freeze the
squeezing dynamics after $\xi^2$ reaches its minimum~\cite{Wu2015}. Indeed, from the aspect
of optimal squeezing angle $\varphi_{\mathrm{opt}}$ shown in Fig.~\ref{fig:angle}, the
learned time-varying control in the unitary dynamics does try to keep $\varphi_{\mathrm{opt}}$
(dash-dotted green line) around the angle to reach the minimum squeezing (dotted black line),
however, when the noise exists, it is difficult for the control to prevent $\varphi_{\mathrm{opt}}$
(dashed blue line) deviating from the angle for $\min \xi^2$ (dotted purple line), and
the combined control can slow down this deviation of $\varphi_{\mathrm{opt}}$ (solid red line)
and thus enlarge the lifetime of squeezing.

All the facts above indicate that the combined controls obtained via our strategy present a
very good performance on both the generation and storage of squeezing. It is not only more
stable and simpler than a full time-varying control, but can also balance the trade-off
between the minimum values of $\xi^2(t)$ and the full amount of the generated squeezing.
Therefore, this strategy could be very helpful in the realistic experiments for the
generation of spin squeezing.

\section{conclusion and outlook}
In conclusion, we have studied the generation and storage of spin squeezing in the collective spin
system coupled to a bosonic field. Three control strategies, constant-value, time-varying, and
combined controls are considered. With a proper constant-value control, this system can simulate
the one- and two-axis twisting models. The time-varying controls show very good performances under
the unitary dynamics, however, when the collective noise is involved, this advantage becomes not
significant.

To deal with this situation, we further propose a strategy for the construction of combined
controls. This strategy contains four steps. The first step is finding an optimal constant-value
control $\zeta$ ($\zeta_{\min}$) that gives the lowest value of $\min \xi^2$, and choose a
reasonable regime $[\zeta_a, \zeta_b]$ around it. The second step is constructing a stitched
control with the constant-value part chosen in $[\zeta_a, \zeta_b]$ and the time-varying part
copied from the full time-varying control. The third step is calculating $S$ for all the stitched
controls and finding the optimal value $\zeta_{\mathrm{opt}}$ which gives the maximum value of $S$.
The fourth step is replacing the time-vary part of the stitched control of $\zeta_{\mathrm{opt}}$
by the one obtained via the learning algorithms, which is just the final combined control. This
combined control is more simple and stable than a full time-varying control. It not only gives
a comparable minimum value of $\xi^2$ with respect to the full time-varying control, but also
provides a better lifetime and larger full amount of squeezing (quantified by $S$). Hence, this
combined strategy is very promising to be applied in practical experiments for a better generation
and storage of spin squeezing.

It is known that the performance of learning might be not good when the action (control) space is
large. Hence, a restriction on the waveform of control field could reduce the action space and help
to find an optimal solution in this space, however, it may also limit the performance of control
since the global optimal control may not be in this chosen subspace. In the meantime, the computational
complexities for the dynamics could dramatically increase when the scale of the system grows. Therefore,
how to make the learning algorithms efficient for a free waveform and how to apply it into large-scale
systems for the sake of enhanced spin squeezing still remains a challenge in this field. More techniques
like the matrix product states may be needed to be involved, and the learning algorithms may also need
to be adjusted correspondingly.

\begin{acknowledgments}
Q.-S.T. acknowledges the support from the NSFC through Grant No. 11805047 and No. 11665010.
J.L. acknowledges the support from NSFC through Grant No. 11805073 and the startup grant
of HUST. J.-Q.L. is supported in part by NSFC (Grants Nos.~11822501, 11774087, and 11935006)
and Hunan Science and Technology Plan Project (Grant No.~2017XK2018).
\end{acknowledgments}

\appendix*
\section{Simulation of one- and two-axis twisting models with constant-value controls}

Here we provide the thorough calculation on the simulation of one-axis twisting and
two-axis twisting type model. Recall the original Hamiltonian is
\begin{equation}
H_0=\omega_{c}a^{\dagger}a+\omega_{z}J_{z}+g J_{x}(a^{\dagger}+a), \label{eq:apx_H0}
\end{equation}
and the control Hamiltonian is
\begin{equation}
H_{\mathrm{c}}(t) = \zeta\nu\cos(\nu t)J_{z}, \label{eq:apx_Hc}
\end{equation}
where $\zeta$ is a time-independent constant.
Notice that the transformation
\begin{eqnarray}
V(t) &:=& \mathcal{T}\exp\left[i\int_{0}^{t}(H_{0}+H_{\mathrm{c}})d\tau\right] \nonumber \\
&=& \exp\left\{i\omega_{c}ta^{\dagger}a+i\left[\omega_{z}t+\zeta\sin(\nu t)\right]J_{z}\right\}
\end{eqnarray}
with $\mathcal{T}$ the time-ordering operator, then in the rotating frame defined by $V(t)$, the
transformed Hamiltonian becomes
\begin{eqnarray*}
& & \tilde{H}(t) \\
&=& V(t)H(t)V^{\dagger}(t)-iV(t)\partial_t{V}^{\dagger}(t) \\
&=& g\left\{\cos\left[\omega_{z}t+\zeta\sin(\nu t)\right]J_{x}
    -\sin\left[\omega_{z}t+\zeta\sin(\nu t)\right]J_{y}\right\} \\
& & \times (a^{\dagger}e^{i\omega_{c}t}+a e^{-i\omega_{c}t}) \\
&=& \frac{g}{2}\left\{e^{i[\omega_{z}t+\zeta\sin(\nu t)]}J_{+}
    +e^{-i[\omega_{z}t+\zeta\sin(\nu t)]}J_{-}\right\} \\
& & \times (a^{\dagger}e^{i\omega_{c}t}+a e^{-i\omega_{c}t})\\
&=& g\mathrm{Re}\!\left(\!e^{i(\omega_{c}+\omega_{z})t}e^{i\zeta\sin(\nu t)}J_{+}a^{\dagger}
    \!+\!e^{-i(\omega_{c}-\omega_{z})t}e^{i\zeta\sin(\nu t)}J_{+}a\right)\!,  \\
\end{eqnarray*}
where $\mathrm{Re}(\cdot)$ is the real part. Utilizing the Jacobi–Anger expansion
\begin{equation}
e^{i\zeta\sin(\nu t)}=\sum^{\infty}_{n=-\infty}\mathcal{J}_{n}(\zeta)e^{in\nu t}
\end{equation}
with $\mathcal{J}_n(\zeta)$ the $n$th Bessel function of the first kind, $\tilde{H}$ can be further rewritten into
\begin{eqnarray*}
\tilde{H} &=& \frac{g}{2}\sum_{m=-\infty}^{\infty}\mathcal{J}_{m}(\zeta)e^{i\Delta_{m}t}J_{+}a^{\dagger} \\
& &  +\frac{g}{2}\sum_{n=-\infty}^{\infty}\mathcal{J}_{n}(\zeta)e^{i\left(n\nu-\delta\right)t}J_{+}a+\mathrm{H.c.},
\end{eqnarray*}
where $\Delta_{m}=m\nu+\omega_{c}+\omega_{z}$, $\delta=\omega_{c}-\omega_{z}$, and H.c. represents the
Hermitian conjugate. Under the condition $\nu\gg g, \delta$, $\tilde{H}$ approximates to
\begin{equation}
\tilde{H}\approx(g_{0}J_{+}a e^{-i\delta t}+g_{m_0}J_{-}a e^{-i\Delta_{m_{0}}t})+\mathrm{H.c.}
\end{equation}
with $g_{0}=g\mathcal{J}_{0}(\zeta)/2$ and $g_{m_0}=g\mathcal{J}_{m_{0}}(\zeta)/2$. Here $m_0$ is an optimal
integer that make $|m\nu+\omega_{\mathrm{c}}+\omega_z|$ minimum. A similar approach~\cite{Huang2017} has
been applied in quantum Rabi model to manipulate the counter-rotating interactions. Now take the condition
$\delta =\Delta_{m_{0}}\equiv\Delta$, that is $m_{0}=-2[\omega_{z}/\nu]$ ($[\cdot]$ is the rounding function),
and transform $\tilde{H}$ back to a non-rotating frame with $V_1=e^{i\Delta a^{\dagger}a t}$, namely,
$\tilde{H}=V_1H_1V_1^{\dagger}-iV_1\partial_t V_1^{\dagger}$, then
\begin{eqnarray}
H_1 &=& \Delta a^{\dagger}a+\frac{g}{2}\left[J_{0}(\zeta)J_{+}+J_{m_{0}}(\zeta)J_{-}\right]a+\mathrm{H.c.}
\nonumber \\
& = & \Delta a^{\dagger}a+g(\Sigma^{\dagger}a+\Sigma a^{\dagger})
\end{eqnarray}
with $\Sigma=[\mathcal{J}_{0}(\zeta)J_{-}+\mathcal{J}_{m_{0}}(\zeta)J_{+}]/2$.

\begin{figure}[tp]
\centering\includegraphics[width=8cm]{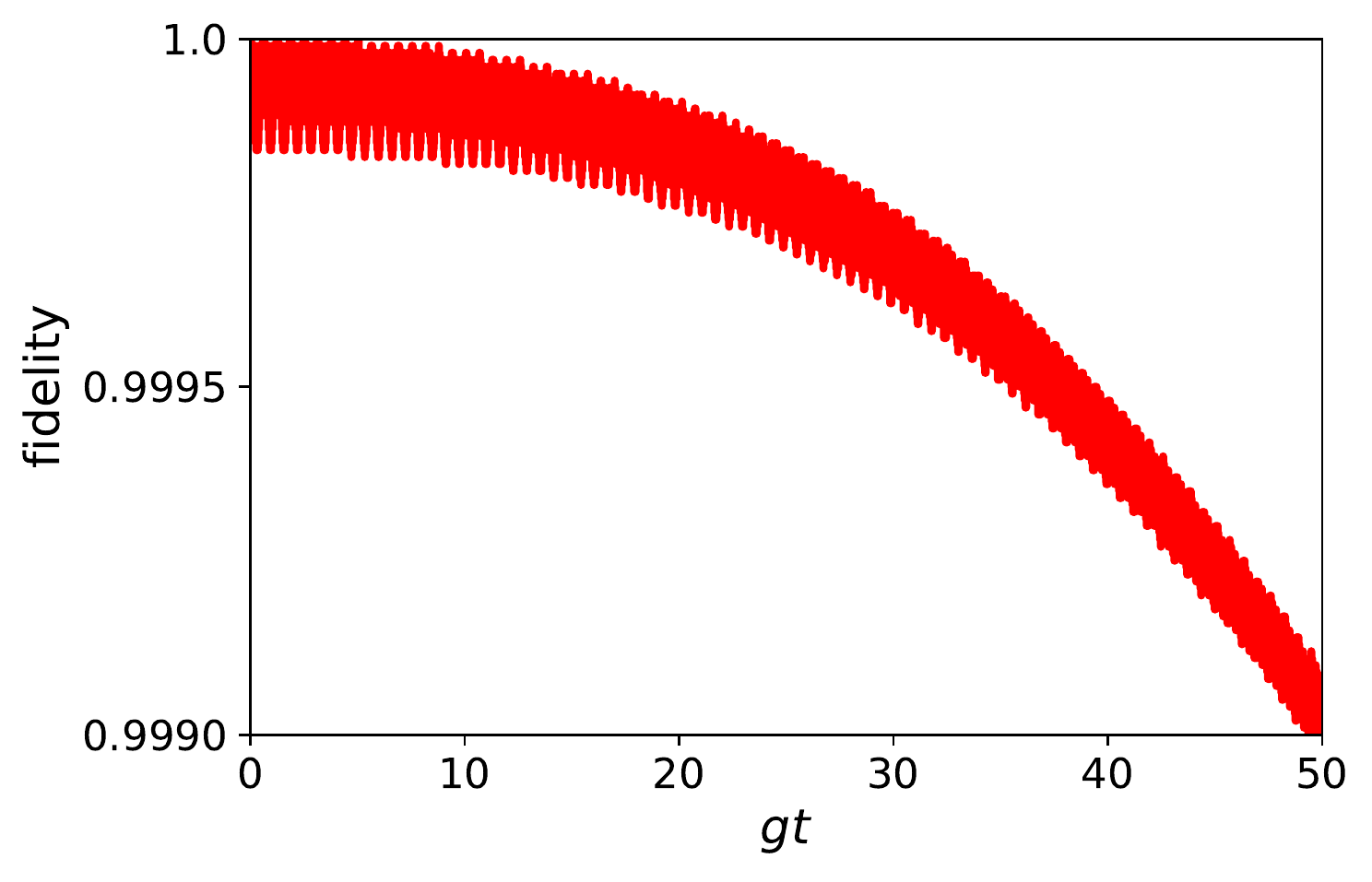}
\caption{The dynamics of the fidelity between the evolved states
given by the original Hamiltonian in Eqs.~(\ref{eq:apx_H0})
and (\ref{eq:apx_Hc}) and the effective Hamiltonian
in Eq.~(\ref{eq:apx_Heff}).}
\label{fig:apx_fidelity}
\end{figure}

In the regime of large detunings $\Delta\gg g > g_{0},g_{m_0}$, $\tilde{H}$ can be approximately
diagonalized by the transformation $e^{R}H_1e^{-R}=H_{\mathrm{eff}}$ with
\begin{equation}
R = \frac{g}{\Delta}(\Sigma a^{\dagger}-\Sigma^{\dagger}a).
\end{equation}
Truncating to the second order of $g/\Delta$, and notice the fact that
\begin{eqnarray}
[R,H_1] &=& -g\Sigma a^{\dagger}-g\Sigma^{\dagger}a -\frac{g^{2}}{\Delta}(\Sigma\Sigma^{\dagger}
    +\Sigma^{\dagger}\Sigma)  \nonumber \\
& & +\frac{g^{2}}{\Delta}[\Sigma, \Sigma^{\dagger}]a a^{\dagger} +\frac{g^{2}}{\Delta}
    [\Sigma, \Sigma^{\dagger}]a^{\dagger}a,
\end{eqnarray}
and
\begin{eqnarray}
[R,[R, H_1]]
&=& \frac{g^2}{\Delta} \left(\Sigma\Sigma^{\dagger}
    +\Sigma^{\dagger}\Sigma\right) - \frac{g^2}{\Delta}
    [\Sigma, \Sigma^{\dagger}]\left(2a^{\dagger}a+1\right)\nonumber \\
&=& -\frac{2g^2}{\Delta} \left([\Sigma, \Sigma^{\dagger}]a^{\dagger}a
    -\Sigma^{\dagger}\Sigma\right),
\end{eqnarray}
$H_\mathrm{{eff}}$ then reduces to
\begin{eqnarray}
H_\mathrm{{eff}} &=& H_1+[R, H_1]+\frac{1}{2}[R, [R, H_1]]+\cdots \nonumber \\
&=& \Delta a^{\dagger}a+\frac{g^2}{\Delta}[\Sigma,\Sigma^{\dagger}]a^{\dagger}a-\frac{g^2}{\Delta}
\Sigma^{\dagger}\Sigma.
\end{eqnarray}
Substituting the expression of $\Sigma$ into the equation above, one can have
\begin{eqnarray}
H_\mathrm{{eff}} &=& \Delta a^{\dagger}a-\left(\frac{g_{0}^{2}-g_{m_0}^{2}}{\Delta}\right)
(1+2a^{\dagger}a)J_{z}  \nonumber \\
& & +\frac{\left(g_{0}-g_{m_0}\right)^{2}}{\Delta}J_{z}^{2}-\frac{4g_{0}g_{m_0}}{\Delta}J_{x}^{2},
\label{eq:apx_Heff}
\end{eqnarray}
in which the constant term has been neglected.
Utilizing this Hamiltonian, an effective one-axis twisting Hamiltonian can be realized by taking
$g_{0}=g_{m_0}$, i.e., $\mathcal{J}_{0}(\zeta)=\mathcal{J}_{m_{0}}(\zeta)$. Under this condition,
$H_{\mathrm{eff}}$ reduces to $-\chi J_{x}^{2}$, with $\chi=\frac{g^{2}}{\Delta}\mathcal{J}_{m_{0}}^{2}
(\zeta)$. In the mean time, the two-axis twisting type Hamiltonian can be obtained by taking
$(g_{0}-g_{m_0})^{2}=4g_{0}g_{m_0}$, which is equivalent to $\mathcal{J}_{0}(\zeta)=
\left(3\pm2\sqrt{2}\right)\mathcal{J}_{m_{0}}(\zeta)$. This condition let $H_{\mathrm{eff}}$
reduce to $-\lambda(J_{x}^{2}-J_{z}^{2})+\lambda^{\prime}(1+2a^{\dagger}a)J_{z}$, where
$\lambda=\frac{4g_0 g_{m_0}}{\Delta}=\frac{g^{2}(3\pm2\sqrt{2})}{\Delta}\mathcal{J}_{m_{0}}^{2}(\zeta)$
and $\lambda^{\prime}=-\frac{g_0^2-g_{m_0}^2}{\Delta}$. The second term in the equation above can be (on average) eliminated through a dynamical decoupling protocol, which makes $H_{\mathrm{eff}}$ a traditional two-axis twisting Hamiltonian $-\lambda(J_{x}^{2}-J_{z}^{2})$.

To show the validity of this effective Hamiltonian,
the fidelity between the evolved states given by the original Hamiltonians (Eqs.~(\ref{eq:apx_H0})
and (\ref{eq:apx_Hc})) and the effective Hamiltonian above are shown in Fig.~\ref{fig:apx_fidelity}
for the unitary dynamics with $N=6$. Other parameters are set as $\omega_z=110$, $\omega_c=100$,
$\nu=200$, $g=1$, and $\zeta=2.569$ (TAT-type). The evolution of fidelity shows that the effective Hamiltonian
works well ($>0.999$) within the time point $gt=50$.

\end{document}